\documentclass[sigconf,screen]{acmart} 
\usepackage{multirow}
\usepackage{subcaption}
\usepackage{hyperref}
\usepackage{fancyhdr} 
\usepackage{tikz}
\usepackage{xcolor}
\usepackage{amsmath}
\usepackage{tabularx}

\usepackage{enumitem}
\usetikzlibrary{positioning, fit, backgrounds, arrows.meta, calc}

\definecolor{contentcolor}{RGB}{255,220,150}
\definecolor{collabcolor}{RGB}{180,210,255}
\definecolor{attentioncolor}{RGB}{255,180,150}
\definecolor{mlpcolor}{RGB}{200,235,200}
\definecolor{grucolor}{RGB}{220,190,255}
\definecolor{artistbg}{RGB}{245,245,245}

\definecolor{mygreen}{RGB}{35, 83, 71}
\definecolor{myred}{RGB}{142, 15, 25}

\usepackage{pifont}
\newcommand{\cmark}{\ding{51}} % ✓
\newcommand{\xmark}{\ding{55}} % ✗

\AtBeginDocument{%
  }

\copyrightyear{2026}
\acmYear{2026}
\setcopyright{cc}
\setcctype{by}
\acmConference[UMAP '26]{34th ACM Conference on User Modeling, Adaptation and Personalization}{June 08--11, 2026}{Gothenburg, Sweden}
\acmBooktitle{34th ACM Conference on User Modeling, Adaptation and Personalization (UMAP '26), June 08--11, 2026, Gothenburg, Sweden}
\acmDOI{10.1145/3774935.3806178}
\acmISBN{979-8-4007-2311-7/2026/06}

\begin{document}

\title{Leveraging Artist Catalogs for Cold-Start Music Recommendation}

% \affiliation{%

\author{Yan-Martin Tamm}
\authornote{Equal contribution}
% \authornote{University of Tartu, Estonia}
\email{yanmart.tamm@gmail.com}
\orcid{0000-0002-6174-7736}
% \affiliation{}
\affiliation{%
  \institution{University of Tartu}
  \streetaddress{Narva mnt 18}
  \city{Tartu}
  \country{Estonia}
  \postcode{51009}
}

\author{Gregor Meehan}
\authornotemark[1]
\authornote{Funded by UKRI CDT in Artificial Intelligence and Music [grant no. EP/S022694/1].}
% \authornote{Queen Mary University of London, United Kingdom}
\email{gregor.meehan@qmul.ac.uk}
\orcid{0009-0007-2619-9299}
% \affiliation{}
\affiliation{%
  \institution{Queen Mary University of London}
  \city{London}
  \country{United Kingdom}
}

\author{Vojtěch Nekl}
% \authornote{Czech Technical University in Prague, Czech Republic}
\email{neklvojt@fit.cvut.cz}
\orcid{0009-0009-2786-5139}
% \affiliation{}
\affiliation{%
  \institution{Czech Technical University in Prague}
  \city{Prague}
  \country{Czech Republic}
}

\author{Vojtěch Vančura}
% \authornotemark[4]
\authornote{Also with Faculty of Mathematics and Physics, Charles University. Supported by the Czech Science Foundation (GAČR) project 25-16785S.}
\email{vojtech.vancura@recombee.com}
\orcid{0000-0003-2638-9969}

\affiliation{%
  \institution{Recombee}
  \city{Prague}
  \country{Czech Republic}
}
% \affiliation{
%     \institution{Faculty of Mathematics and Physics, Charles University}
%     \city{Prague}
%   \country{Czech Republic}
% }
\author{Rodrigo	Alves}
% \authornotemark[4]
\email{rodrigo.alves@fit.cvut.cz}
\orcid{0000-0001-7458-5281}
% \affiliation{}
\affiliation{%
  \institution{Czech Technical University in Prague}
  \city{Prague}
  \country{Czech Republic}
}

\author{Johan Pauwels}
% \authornotemark[3]
\email{j.pauwels@qmul.ac.uk}
\orcid{0000-0002-5805-7144}
% \affiliation{}
\affiliation{%
  \institution{Queen Mary University of London}
  \city{London}
  \country{United Kingdom}
}

\author{Anna Aljanaki}
% \authornotemark[2]
\email{aljanaki@gmail.com}
\orcid{0000-0002-7119-8312}
% \affiliation{}
\affiliation{%
  \institution{University of Tartu}
  \streetaddress{Narva mnt 18}
  \city{Tartu}
  \country{Estonia}
  \postcode{51009}
}
% \thanks{\textsuperscript{1}Institute A \quad \textsuperscript{2}Institute B}

\renewcommand{\shortauthors}{Tamm and Meehan et al.}

\begin{abstract}
The item cold-start problem poses a fundamental challenge for music recommendation: newly added tracks lack the interaction history that collaborative filtering (CF) requires. Existing approaches often address this problem by learning mappings from content features such as audio, text, and metadata to the CF latent space. However, previous works either omit artist information or treat it as just another input modality, missing the fundamental hierarchy of artists and items. Since most new tracks come from artists with previous history available,  we frame cold-start track recommendation as `semi-cold' by leveraging the rich collaborative signal that exists at the artist level. We show that artist-aware methods can more than double Recall and NDCG compared to content-only baselines, and propose ACARec, an attention-based architecture that generates CF embeddings for new tracks by attending over the artist's existing catalog. We show that our approach has notable advantages in predicting user preferences for new tracks, especially for new artist discovery and more accurate estimation of cold item popularity.
\end{abstract}


% \ccsdesc[500]{Information systems~Music retrieval}
% \ccsdesc[500]{Computing methodologies~Neural networks}

% \keywords{music recommender systems, recommender systems, cold start, hybrid recommender systems}

% \received{20 February 20077}

\hyphenation{MusicFM MERT EncodecMAE Jukebox MusiCNN HitRate}


\maketitle

% Things that move us towards a long paper:
% \begin{itemize}
%     \item complex study
%     \item thorough discussion of related work
%     \item sufficient experimental validation
%     \item strong theoretical foundation
% \end{itemize}

\section{Introduction}

The cold-start problem~\cite{schein2002methods} remains one of the most persistent challenges in music recommendation systems (MRSs)~\cite{Content_driven_music_rec}. Streaming platforms add new releases continuously, and the ability to surface fresh tracks to the right listeners directly affects user satisfaction~\cite{ungruh2024putting,ferwerda2023don}. However, when a new track enters the catalog, it has no interaction data, the primary signal that collaborative filtering (CF) methods depend on, while users still expect immediately relevant recommendations~\cite{zhang2025cold}.

This problem is typically addressed by learning a mapping from content features to CF latent spaces. DeepMusic \cite{van2013deep} pioneered this approach for music, training CNNs on audio spectrograms to predict latent factors for tracks with no listening history. Many refinements followed: joint training of content and collaborative objectives~\cite{liang2015content, NCACF}, contrastive alignment instead of regression~\cite{CLCREC}, and similarity-based retrieval formulations~\cite{pulis2021siamese}.

However, as observed by \cite{van2013deep}, audio alone cannot capture all aspects of user preference, which motivated subsequent work to incorporate additional modalities and metadata besides track content. For instance, prior work has incorporated artist information, such as artist biographies as additional text inputs~\cite{oramas2017deep}, or artist embeddings as an additional input into the user tower~\cite{chen2021learning};  more recently, LARP~\cite{salganik2024larp} includes artist names in generated text descriptions. 

We note that these studies typically use artist information \emph{only} as an auxiliary feature rather than a primary signal. Recent work~\cite{meehan2025artist}, in contrast,  demonstrates that even a simple heuristic of recommending a user random tracks from artists they have previously listened to can compete with (and often beat) elaborate track-based cold-start methods. This result indicates that artist identity carries a collaborative signal that content-based methods fail to capture, motivating a deeper investigation into how artist information can be exploited for cold-start recommendation. Treating each track in isolation and relying solely on its acoustic or metadata features to infer listener preferences overlooks a crucial aspect of artist-track mapping: the vast majority of new tracks are released by artists already represented in the training data. For these tracks, the cold-start problem is not truly cold: it possesses rich collaborative signal about how listeners respond to this artist's existing work.

%Some prior work has incorporated artist information, but typically as an auxiliary feature rather than a primary signal. \cite{oramas2017deep} uses artist biographies as additional text input, \cite{chen2021learning} uses artist embeddings as an additional input into the user tower and even recent approaches like LARP~\cite{salganik2024larp} mention artist names only in generated text descriptions rather than leveraging artist identity directly. The potential of artist-track mapping remains underexplored.

Based on these insights, in this paper we reframe cold-start track embedding as a `semi-cold' problem over the context of the artist's existing catalog. Given a new track's content features, we ask: which existing tracks by that artist are most informative for predicting listener interest? This perspective naturally leads to an attention-based architecture that learns to aggregate relevant artist catalog tracks based on similarity to the new release. This \textbf{A}rtist \textbf{C}atalog \textbf{A}ttention (\textbf{ACARec})\footnote{To support reproducibility of our work, we make our code publicly available at \href{https://github.com/gmeehan96/ACARec}{https://github.com/gmeehan96/ACARec}} approach substantially outperforms both content-only baselines and their artist-aware modifications, as well as naive catalog aggregation methods, showing that the artist catalog provides a powerful bridge across the cold-start semantic gap. Moreover, we adopt an artist-aware evaluation methodology~\cite{meehan2025artist} and investigate ACARec's cold-start performance across new-artist discovery and known-artist exploitation scenarios.

\section{Related Work}

\subsection{Item Cold-Start}
In this study, we consider cold-start methods leveraging item multimedia and metadata features for direct inference on new items. We acknowledge that the use of Large Language Models (LLMs) is an emerging paradigm in cold item recommendation~\cite{zhang2025cold, huang2025large}; although promising, these methods face scalability challenges for widespread industry deployment~\cite{wang2024fresh}. We therefore focus on lightweight content-based methods with low latency and inference costs. 

\subsubsection{Content-Based Cold-Start} Content features, such as images, text descriptions, audio content, or other metadata, are often available for new items and therefore are a valuable resource for predicting cold item preferences. Dropout-based methods, such as DropoutNet~\cite{volkovs2017dropoutnet}, train a hot model to make cold item predictions by randomly swapping between CF and content embeddings during training, simulating the cold scenario. More recent works extend this approach with improved content projections (e.g.\ mixtures-of-experts~\cite{HEATER}), graph neighborhoods~\cite{kim2024content}, or contrastive training objectives to align content and CF embeddings~\cite{CLCREC,monteil2024marec,zhou2023contrastive,wang2024preference}.

However, training for prediction on both cold and warm items in a single model can damage warm item accuracy~\cite{huang_aligning_2023}. Many methods, therefore, first train a CF model then teach a content encoder to project into its embedding space, e.g., with reconstruction loss~\cite{van2013deep} or variational autoencoders~\cite{zhao2022improving, bai2023gorec}. Others~\cite{huang_aligning_2023,chen2022generative,sun2020lara} train cold item models to replicate CF model ranking behavior. 

A limitation of these content-based solutions is the information gap between item content and collaborative signal. For example, recent work~\cite{meehan2025inherited} shows that cold-start models imitate popularity bias~\cite{klimashevskaia2024survey} in the supervisory CF system; however, these models are predicting cold item popularity based only on their content features, leading to some items being suggested far beyond their actual level of interest. Our proposed ACARec method alleviates this problem in music contexts by recognizing that artist catalogs provide a rich source of CF signal for new tracks, narrowing this information gap.

\subsubsection{Cold-Start in MRSs} Cold-start MRS works can be divided into two categories. Purely content-based approaches~\cite{mcfee2012learning,salganik2024larp, borges2023audio, alonso-jimenez_pre-training_2023, ferraro2021enriched, meehan2025evaluating, pulis2021siamese} focus on learning improved representations of musical audio for cold-start MRSs. Their aim is for similarity in the learned space to correlate with user interests; this is typically accomplished with supervision from user interaction histories, e.g.\ contrastive pairs of songs from the same playlist~\cite{meehan2025evaluating,ferraro2021enriched,alonso-jimenez_pre-training_2023,salganik2024larp}. The second category contains hybrid approaches such as DeepMusic~\cite{van2013deep} and NCACF~\cite{NCACF}. These and other methods~\cite{ganhor2024multimodal, oramas2017deep} are hybrid because they supplement the content-based training process with CF embeddings. ACARec fits in this hybrid category, as we use a pre-trained audio encoder and leverage a separate CF model for supervision.

All of these studies use audio features as a primary source of cold track content. SiBraR~\cite{ganhor2024multimodal} also includes other data modes, such as genre tags and album images. Many MRS works also leverage artist information, as we discuss in Section \ref{sec:related_artist}; however, to our knowledge, existing methods do not consider how CF signal from an artist's previous tracks can be exploited for new tracks.

\subsection{Artist-Aware Music Recommendation}\label{sec:related_artist}
In this paper, we focus on how artist metadata is used for track-level MRSs, rather than for the separate tasks of artist-level recommendation\ \cite{mcfee2012learning, bertram2023all, trainor2023popularity} or artist similarity\ \cite{grotschla2024towards,oramas2024talking,Korzeniowski-2022,da2024artist}. The pioneering multi-modal track-level approach that included artist information \cite{oramas2017deep} combines artist biographies, audio features, and user feedback to address the cold-start problem, aggregating artist catalogs for improved artist embeddings and recommendation quality.

Subsequent work explores additional modalities beyond artist biographies. One study \cite{chen2021learning} proposes a Siamese metric learning approach, where the item tower processes track spectrograms while the user tower combines demographic features with embeddings of recent tracks, albums, and artists. Another \cite{briand2024let} describes a production system at Deezer that predicts embeddings for new albums using various metadata, including the artist, following the method from \cite{van2013deep} but operating at album-level rather than track-level.

Another line of research explores how artist metadata can be leveraged to learn improved representations of musical audio for downstream MRS tasks. Some methods \cite{alonso-jimenez_pre-training_2023,meehan2025evaluating} construct contrastive pairs based on artist identities, while others represent heterogeneous metadata as natural language captions and process them through a pretrained text encoder \cite{salganik2024larp,Lee2025} such as BERT \cite{devlin2019bert}. 

% One such approach \cite{salganik2024larp} proposes a multi-modal cold-start playlist continuation model employing a contrastive framework at three levels: within-track, track-track, and track-playlist. At each level, they calculate the contrast between the audio and textual representations. The textual representation is a caption constructed as "The track \texttt{<track name>} by \texttt{<artist name>} on album \texttt{<album name>}". This approach captures not only track, artist, and album identities but also the semantic content of their names. However, it remains unclear how much of the improvement stems from linking tracks through shared artists, which could also be achieved with artist IDs rather than encoding the semantic meaning of artist names.

Semantic IDs \cite{rajput2023recommender,singh2024better} have recently emerged as a promising direction in recommender systems and have been applied in MRSs~\cite{Lee2025,mei2025semantic}. Rather than treating item IDs as arbitrary indices, this approach replaces them with learned sequences of meaningful discrete codes (typically generated by RQ-VAEs \cite{zeghidour2021soundstream}) that capture semantic similarity and enable better generalization to new items. For example, \cite{Lee2025} applies a multimodal music tokenizer to textual descriptions (e.g.\ artist names, band members, song facts, artist biographies) with a pretrained text encoder for generative retrieval. Another method \cite{mei2025semantic} constructs item embeddings as a sum of track, artist, and genre embeddings; while \cite{mei2025semantic} primarily focuses on the resulting semantic IDs, this composition model outperforms the version using only track embeddings, hinting at the importance of artist information. Moreover, their artist-enriched embeddings allow randomly initialized semantic IDs to match the performance of trained ones, which the authors note might be because "artist and genre embeddings perform a similar function to the semantic IDs".

The above work highlights the crucial role of artist information in modeling user preferences. However, existing methods typically treat artists as just another metadata feature, rather than studying their particular impact. Graph-based methods \cite{liu2024music,bevec2024hybrid,MKGCN,GASM,wang2023multi} that treat artists as nodes similarly use them as auxiliary input alongside other metadata. We argue that the artist's role is more fundamental, and that artist catalogs should be incorporated directly into the model, prior to and distinct from text captions or biographies.

% \subsection{Other similar stuff}
% \begin{itemize}
%     \item Alleviating Cold Start in the EOSC Recommendations: Extended Page Rank Algorithm  
%     \begin{itemize}
%         \item adds co-authourship information to improve user-side cold start recommendation
%     \end{itemize}
%     \item Simple and effective neural-free soft-cluster embeddings for item cold-start recommendations
%     \begin{itemize}
%         \item Use item side information to help with clustering
%     \end{itemize}
% \end{itemize}
 
% \newpage
\section{ACARec}

\paragraph{\textbf{Notation}} Let $\mathcal{U}$ and $\mathcal{I}$ denote sets of users and items respectively. We partition items into \emph{hot} items $\mathcal{H}$, whose interactions are observed in the training data, and \emph{cold} items $\mathcal{C}$, which first appear in the test data, with $\mathcal{I}=\mathcal{H}\cup\mathcal{C}$ and $\mathcal{H}\cap\mathcal{C}=\emptyset$. We represent implicit feedback with a binary relevance matrix $\mathbf{R}\in\{0,1\}^{|\mathcal{U}|\times |\mathcal{I}|}$, where $R_{u,i}=1$ if user $u$ interacted with item $i$ and $0$ otherwise. By construction, for any $c\in\mathcal{C}$ we have $R_{u,c}=0$ for all $u\in\mathcal{U}$.

Each item $i\in\mathcal{I}$ is represented by an audio content embedding $\mathbf{x}_i\in\mathbb{R}^{d_c}$. Stacking these embeddings yields a content matrix $\mathbf{X}\in\mathbb{R}^{|\mathcal{I}|\times d_c}$. Each item $i$ also has a corresponding artist $a_i$; we discuss the assumption that each track only has a single artist in Section \ref{sec:limitations}. We let $\mathcal{H}_a=\{h\in\mathcal{H}:a_h=a\}$ be the set of hot items by artist $a$, and write $\mathbf{X}_a\in\mathbb{R}^{|\mathcal{H}_a|\times d_c}$ for the submatrix of $\mathbf{X}$ indexed by $\mathcal{H}_a$.

We also assume access to a pre-trained CF model trained on the observed interactions $\mathbf{R}$.
%, obtained by solving a generic objective of the form
%\begin{equation}
%    \min_{\{\mathbf{p}_u\}_{u\in\mathcal{U}},\,\{\mathbf{e}_h\}_{h\in\mathcal{H}}}
%    \; \mathcal{L}_{\mathrm{CF}}\!\left(\mathbf{R}; \{\mathbf{p}_u\}, \{\mathbf{e}_h\}\right).
%\end{equation}
The CF model assigns latent embeddings $\mathbf{p}_u\in\mathbb{R}^{d_e}$ to each user $u\in\mathcal{U}$ and $\mathbf{e}_h\in\mathbb{R}^{d_e}$ to each hot item $h\in\mathcal{H}$; we stack these into user and item embedding matrices $\mathbf{P}\in\mathbb{R}^{|\mathcal{U}|\times d_e}$ and $\mathbf{E}\in\mathbb{R}^{|\mathcal{H}|\times d_e}$, respectively. As for $\mathbf{X}_a$, for artist $a$ we write $\mathbf{E}_a\in\mathbb{R}^{|\mathcal{H}_a|\times d_e}$ for the submatrix of $\mathbf{E}$ containing the CF embeddings of the hot items by $a$. User-item preference scores for a $({u,h})$ pair are predicted by dot products $\mathbf{p}_u^\top \mathbf{e}_h$.

\subsection{Problem Definition}  %Our approach is motivated by two observations. First, users’ listening histories are often concentrated within a small set of artists, and these artists are frequently closely related. Since an artist’s catalog tends to share musical characteristics, artist identity provides a particularly informative form of side information and often correlates with attributes such as genre, instrumentation, and production style. Second, even when a newly released track is cold and has no interaction history, its artist is typically not cold: the artist usually has existing hot tracks with observed interactions, from which collaborative representations can be learned.

To infer CF embeddings ${\mathbf{e}}_c\in\mathbb{R}^{d_e}$ for cold items $c\in\mathcal{C}$, prior works typically learn a mapping of \emph{item-specific} content ($\mathbf{x}_c$) into CF-space: 
%Existing methods accomplish this by training a predictor that maps item content into the collaborative embedding space:
\begin{equation}
    \hat{\mathbf{e}}_c = f_{\theta}\!\left(\mathbf{x}_c\right),
\end{equation}
where $\theta$ denotes latent parameters of $f_{\theta}$, which are learned by using hot items as supervision. Then, at inference, $\hat{\mathbf{e}}_c$ can be used to predict cold-item preference scores (${\hat{\mathbf{e}}_c}^\top, \mathbf{p}_u$) for user $u$.

Although a cold item $c \in \mathcal{C}$ has no learned CF embedding $\mathbf{e}_c$, it has an artist $a_c$, which will typically have hot items (i.e, $|\mathcal{H}_{a_c}|>0$). The embeddings for these items provide indirect warm content and collaborative signals ($\mathbf{X}_{a_c},\mathbf{E}_{a_c}$) that can be leveraged to construct an improved embedding for $c$. We therefore define 
%However, although cold items $c\in\mathcal{C}$ have no learned CF embedding $\mathbf{e}_c$, we observe that their artist may have hot items; when $|\mathcal{H}_{a_c}|>0$, the matrix $\mathbf{E}_{a_c}$ provides collaborative information that can be leveraged to construct an improved embedding for $c$.  We therefore define an augmented predictor as
\begin{equation}
    \hat{\mathbf{e}}_c = g_{\theta}\!\left(\mathbf{x}_c,\mathbf{X}_{a_c},\mathbf{E}_{a_c}\right),
\end{equation}
%also including the content matrix $\mathbf{X}_{a_c}$ 
as an augmented predictor to capture similarities between $c$ and existing tracks by $a_c$. Similarly to DeepMusic~\cite{van2013deep}, we train $g_{\theta}$ to minimize a reconstruction loss on the hot items $h$:
\begin{equation}
    \mathcal{L}_\theta=\sum_{h\in\mathcal{H}} \left\|g_\theta\!\left(\mathbf{x}_h,\mathbf{X}_{a_h},\mathbf{E}_{a_h}\right)-\mathbf{e}_h\right\|^2.
\end{equation}
\paragraph{\textbf{Remark: }}\label{par:remark} Although the items used as supervision during training are hot, we mimic the cold-start setting by withholding the target item from its artist context when forming $(\mathbf{X}_{a_h},\mathbf{E}_{a_h})$. For each target $h\in\mathcal{H}$ with artist $a_h$, we sample a fixed-size context set $\mathcal{H}'_{a_h}\subseteq \mathcal{H}_{a_h}\setminus\{h\}$ and construct $\mathbf{X}'_{a_h}$ and $\mathbf{E}'_{a_h}$ using only items in $\mathcal{H}'_{a_h}$. Sampling a fixed number of context items reduces sensitivity to variation in artist catalog size and keeps training batches computationally stable. At inference time, for a cold item $c$ we use the full hot item set $\mathcal{H}_{a_c}$ by default, although we explore variations on this approach in Section \ref{sec:sampling_results}. For notational simplicity, we write $\mathcal{H}_a$ to denote the artist context set in both regimes (interpreted as the sampled set $\mathcal{H}'_a$ during training and the full set $\mathcal{H}_a$ at inference), and similarly use $\mathbf{X}_a$ and $\mathbf{E}_a$ for the corresponding context matrices.

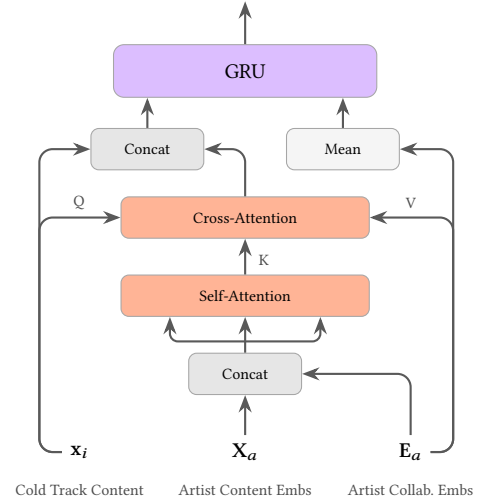
\begin{figure}
    \begin{tikzpicture}[
    >=Stealth,
    block/.style={
        rectangle, 
        draw=gray!70, 
        rounded corners=4pt, 
        minimum width=2.2cm, 
        minimum height=0.7cm, 
        align=center, 
        font=\small
    },
    wideblock/.style={
        rectangle, 
        draw=gray!70, 
        rounded corners=4pt, 
        minimum width=3.5cm, 
        minimum height=0.7cm, 
        align=center, 
        font=\small
    },
    smallblock/.style={
        rectangle, 
        draw=gray!70, 
        rounded corners=3pt, 
        minimum width=1.5cm, 
        minimum height=0.55cm, 
        align=center, 
        font=\scriptsize
    },
    tinyblock/.style={
        rectangle, 
        draw=gray!70, 
        rounded corners=2pt, 
        minimum width=0.9cm, 
        minimum height=0.45cm, 
        align=center, 
        font=\scriptsize
    },
    attnbox/.style={
        rectangle, 
        draw=gray!70, 
        rounded corners=3pt, 
        minimum width=3.3cm, 
        minimum height=0.55cm, 
        align=center, 
        font=\scriptsize,
        fill=attentioncolor
    },
    attentionframe/.style={
        rectangle,
        draw=gray!60,
        dashed,
        rounded corners=8pt,
        fill=none,
        inner sep=8pt
    },
    inputstyle/.style={font=\small},
    outputstyle/.style={font=\small\bfseries},
    annotstyle/.style={font=\scriptsize, text=gray!60!black},
    arrow/.style={->, thick, gray!70!black},
    dotmark/.style={circle, fill=gray!70!black, inner sep=1.5pt}
]

% =====================
% COLUMN POSITIONS
% =====================

\pgfmathsetmacro{\colWidth}{2.2}
\pgfmathsetmacro{\colA}{0}
\pgfmathsetmacro{\colB}{\colA + \colWidth}
\pgfmathsetmacro{\colC}{\colB + \colWidth}
\pgfmathsetmacro{\colAB}{(\colA + \colB) / 2}
\pgfmathsetmacro{\colBC}{(\colB + \colC) / 2}

% W block spacing
\pgfmathsetmacro{\wSpacing}{1.3}
\pgfmathsetmacro{\vSpacing}{1.3}

% =====================
% ROW 0: INPUTS
% =====================

\node[inputstyle] (xi) at (\colA, 0) {$\mathbf{x}_i$};
\node[inputstyle] (XA) at (\colB, 0) {$\mathbf{X}_a$};
\node[inputstyle] (EA) at (\colC, 0) {$\mathbf{E}_a$};

\node[annotstyle] at (\colA, -0.5) {Cold Track Content};
\node[annotstyle] at (\colB, -0.5) {Artist Content Embs};
\node[annotstyle] at (\colC, -0.5) {Artist Collab. Embs};

% =====================
% ROW 1: CONCAT
% =====================
\node[smallblock, fill=gray!20] (concat1) at (\colB, 0.8*\vSpacing) {Concat};
% \node[annotstyle] at ($(concat1.west) + (-0.3, 0)$) {${\mathbf{Y}}_A$};

\node[attnbox] (sa_attn) at (\colB, 1.6*\vSpacing) {Self-Attention};
% Main line from lin_ya going up to W_K, with splits to W_Q and W_V
% \draw[arrow] (lin_ya.north) -- (sa_wk.south);
% Split point
\coordinate (trident_split) at ($(concat1.north) + (0, 0.15)$);
\draw[arrow, rounded corners=3pt] (trident_split) -| ([xshift=-1cm]sa_attn.south);
\draw[arrow, rounded corners=3pt] (trident_split) -| ([xshift=1cm]sa_attn.south);
% \node[annotstyle] at ($(trident_split) + (0.3, 0.2)$) {${\mathbf{Y}}_A$};

% Output label
% \node[annotstyle] at ($(sa_attn.north) + (0.25, 0.2)$) {$\widetilde{\mathbf{E}}_A$};
\node[annotstyle] at ($(sa_attn.north) + (0.25, 0.2)$) {K};

% =====================
% CROSS-ATTENTION BLOCK (expanded)
% =====================

% Background frame for cross-attention (dotted)
\node[attnbox] (ca_attn) at (\colB, 2.4*\vSpacing) {Cross-Attention};
% \node[annotstyle] at ($(ca_attn.east) + (0.3, 0.3)$) {Eq.~5};
\node[annotstyle] at ($(ca_attn.east) + (0.55, 0.2)$) {V};
\node[annotstyle] at ($(ca_attn.west) + (-0.55, 0.2)$) {Q};

% W_Q, W_K, W_V inside cross-attention

% \node[annotstyle] at ($(ca_wq.north) + (0.2, 0.2)$) {$Q$};
% \node[annotstyle] at ($(ca_wk.north) + (0.2, 0.2)$) {$K$};
% \node[annotstyle] at ($(ca_wv.north) + (0.2, 0.2)$) {$V$};

% Attention(Q,K,V) inside cross-attention (orange, wider)

% Output label
% \node[annotstyle] at ($(ca_attn.north) + (0.2, 0.15)$) {$\dot{\mathbf{e}}_i$};

% =====================
% CONCAT for output (aligned with Linear below)
% =====================

% \node[smallblock, fill=gray!20] (concat2) at (\colB - \wSpacing, 12) {Concat};

% =====================
% LINEAR and MEAN under GRU (positioned for straight arrows)
% =====================

\node[smallblock, fill=gray!20] (concat2) at (\colB - \wSpacing, 3.1*\vSpacing) {Concat};
% \node[annotstyle, above left=-0.1cm and -0.1cm of concat2] {Eq.~6};

\node[smallblock, fill=artistbg, draw=gray!70] (mean) at (\colB + \wSpacing, 3.1*\vSpacing) {Mean};
% \node[annotstyle, above right=-0.1cm and -0.1cm of mean] {Eq.~8};

% =====================
% GRU (wide block)
% =====================

\node[wideblock, fill=grucolor] (gru) at (\colB, 3.9*\vSpacing) {GRU};
% \node[annotstyle, right=0.0cm of gru] {Eq.~10};

% =====================
% OUTPUT
% =====================

% \node[outputstyle] (output) at (\colB, 4.8*\vSpacing) {$\hat{\mathbf{e}}_i^{\mathrm{GRU}}$};
\node[outputstyle] (output) at (\colB, 4.7*\vSpacing) {};

% =====================
% ARROWS: Inputs to first processing
% =====================

% X_A to Concat
\draw[arrow] (XA) -- (concat1);

% E_A to Concat and to Linear
\draw[arrow, rounded corners=5pt] (EA) |- (concat1.east);

\draw[arrow, rounded corners=5pt] (EA.east) -- ++(0.3, 0) |- (mean.east);
\draw[arrow, rounded corners=5pt] (EA.east) -- ++(0.3, 0) |- (ca_attn.east);
\draw[arrow, rounded corners=5pt] (xi.west) -- ++(-0.3, 0) |- (concat2.west);
\draw[arrow, rounded corners=5pt] (xi.west) -- ++(-0.3, 0) |- (ca_attn.west);

% =====================
% ARROWS: Concat to Linear Y_A
% =====================
\draw[arrow] (concat1) -- (sa_attn.south);

% =====================
% ARROWS: To Self-Attention inputs (trident pattern)
% =====================

% =====================
% ARROWS: To Cross-Attention inputs (fixed routing - arrive from bottom, lower turns)
% =====================

% Self-attention output (K) to cross-attention W_K
\draw[arrow] (sa_attn.north) -- (ca_attn.south);
\draw[arrow, rounded corners=5pt] (ca_attn.north) |- (concat2.east);

% =====================
% ARROWS: To GRU (straight up)
% =====================

% \draw[arrow] (concat2) -- node[annotstyle, pos=0.5, left] {$\widetilde{\mathbf{e}}_i$} ($(gru.south) + (-\wSpacing, 0)$);
\draw[arrow] (concat2) -- ($(gru.south) + (-\wSpacing, 0)$);

% \draw[arrow] (mean) -- node[annotstyle, pos=0.5, right] {$\overline{\mathbf{e}}_A$} ($(gru.south) + (\wSpacing, 0)$);
\draw[arrow] (mean) -- ($(gru.south) + (\wSpacing, 0)$);

% =====================
% ARROW: GRU to output
% =====================

\draw[arrow] (gru) -- (output);

\end{tikzpicture}
    \caption{ACARec model architecture. The attention blocks include the standard input and output linear projections.}
    \label{fig:model}
\end{figure}

\subsection{Model Architecture}
We now specify the predictor $g_\theta(\cdot)$ for a target item $t$ (a pseudo-cold hot item during training, or a cold item at inference) with artist $a=a_t$. We display its architecture in Figure \ref{fig:model}. 
\subsubsection{Artist Catalog Attention}
We first define the concatenation of the context matrices $\mathbf{E}_a$ and $\mathbf{X}_a$ as
\begin{equation}
    \mathbf{Y}_a = [\mathbf{X}_a;\mathbf{E}_a] \in \mathbb{R}^{|\mathcal{H}_a| \times (d_c+d_e)},
\end{equation}
then contextualize the artist catalog via self-attention:
\begin{equation}\label{eq:self-attn}
    \widetilde{\mathbf{Y}}_a = \mathrm{MH}(\mathbf{Y}_a,\mathbf{Y}_a,\mathbf{Y}_a),
\end{equation}
where $\mathrm{MH}(Q,K,V)$ is a multi-head attention block ~\cite{vaswani2017attention} (including input and output linear projections) with queries $Q$, keys $K$, and values $V$. Then, we compute a content-conditioned summary of the artist’s collaborative embeddings using cross-attention, with the target content as the query, the contextualized catalog as keys, and the artist collaborative embeddings as values:
\begin{equation}
    \dot{\mathbf{e}}_t = \mathrm{MH}(\mathbf{x}_t,\widetilde{\mathbf{Y}}_a,\mathbf{E}_a).
\end{equation}
Finally, we concatenate the attention output and content input, so that the target item's content directly influences the reconstruction:
\begin{equation}\label{eq:cont_inp}
    \widetilde{\mathbf{e}}_t = [\dot{\mathbf{e}}_t;\mathbf{x}_t].
\end{equation}

\subsubsection{Residual Fusion}\label{sec:resid_fusion}
While $\widetilde{\mathbf{e}}_t$ provides a content-conditioned summary of artist context, an artist's CF item embeddings are often centered around a shared artist-specific vector. This suggests using an artist-level prototype as a stable anchor for prediction, and learning deviations from this anchor specific to the target track. We use the mean CF embedding of the artist context as this prototype:
\begin{equation}\label{eq:artist_mean}
    \overline{\mathbf{e}}_{a}=\frac{1}{|\mathcal{H}_a|}\sum_{j\in\mathcal{H}_a}\mathbf{e}_j.
\end{equation}
The target embedding can then be predicted as an additive residual around this mean:
\begin{equation}
    \hat{\mathbf{e}}_t^{\mathrm{Resid}}=\overline{\mathbf{e}}_{a}+(\mathbf{W}\widetilde{\mathbf{e}}_t+\mathbf{b}),
\end{equation}
where $\mathbf{W}\in\mathbb{R}^{d_e\times(d_h+d_c)}$ and $\mathbf{b}\in\mathbb{R}^{d_e}$. This parameterization biases the model toward producing embeddings that remain in the artist’s collaborative neighborhood, while allowing content-dependent corrections when the target track deviates from the artist average.

\subsubsection{Learnable Fusion}
A fixed additive residual may still be too rigid, since the relevance of the artist mean can vary across tracks and artists (e.g., artists with heterogeneous catalogs, or tracks whose audio is atypical for the artist). To allow the model to adaptively trade off between the artist prototype and the content-conditioned signal, we combine them with a gating mechanism using a single update from a Gated Recurrent Unit (GRU) \cite{cho2014properties}:
\begin{equation}
    \hat{\mathbf{e}}_t^{\mathrm{GRU}}=\mathrm{GRU}\!\left(\overline{\mathbf{e}}_{a},\,\widetilde{\mathbf{e}}_t\right).
\end{equation}
Here, the GRU’s update gate controls how strongly $\widetilde{\mathbf{e}}_t$ modifies the artist mean, enabling item-specific interpolation between the artist prototype and attention-based prediction. We use this GRU strategy by default, and evaluate other fusion mechanisms in Section \ref{sec:ablation}.

%At each timestep $t$, a GRU produces its new hidden state as a linear interpolation between the hidden state from the previous timestep ($\mathbf{h}_{t-1}$) and a new candidate state conditioned on the input at timestep $t$ ($\mathbf{x}_{t}$), i.e.\
%\begin{equation}
%    \mathrm{GRU}(\mathbf{h}_{t-1},\mathbf{x}_{t}) = (1-\mathbf{z}_t) %\odot \mathbf{h}_{t-1} + \mathbf{z}_t\odot\widetilde{\mathbf{h}}_{t}
%\end{equation}
%where $\odot$ is elementwise multiplication and both $\mathbf{z}_t$ (the update gate) and $\widetilde{\mathbf{h}}_t$ (the candidate state) are MLP-based functions of $\mathbf{h}_{t-1}$ and $\mathbf{x}_t$.

%We saw in Section Y.Y that cold-start performance can be improved by leveraging artist information for newly added tracks, even by simply incorporating the artist mean collaborative embedding. However, a simple mean produces the same embedding for all tracks of the artist regardless of the track contents, which becomes more problematic the more tracks an artist has as it smooths the diversity in the catalog. 

\section{Experimental Setup}
We design our experiments to address five research questions:
\begin{itemize}[leftmargin=*]
    \item \textbf{RQ1}: To what extent are content-based cold-start methods improved by augmentation with artist context?
    \item \textbf{RQ2}: How do ACARec's cold-start results compare to artist-aware baselines across Overall, Discovery, and Exploit settings?
    \item \textbf{RQ3}: How does ACARec differ from existing methods in its predictive behavior, especially in terms of item popularity?
    \item \textbf{RQ4}: What is the effect of artist sampling strategies on ACARec?
    \item \textbf{RQ5}: What is the impact of ablating key components of ACARec?
\end{itemize}

\subsection{Evaluation}
\begin{table}
\small
\centering
\caption{Train-test split statistics}
\label{tab:data}
    \begin{tabular}{lccccc}
    \toprule
    \multicolumn{6}{c}{\textbf{M4A-Onion}} \\
    \midrule
    & Train & Val & Test & Discovery & Exploit \\
    \midrule
    Interactions    & 5,285,859 & 58,552 & 108,618 & 51,769 & 56,849 \\
    Users   & 20,925    & 8,205  & 8,826   & 7,086 & 6,581 \\
    Items   & 50,249    & 466    & 1,474   & 1,422 & 1,413 \\
    Artists & 9,980     & 263    & 632     & 611   & 597 \\
    \midrule
    \multicolumn{6}{c}{\textbf{Yambda-50m}} \\
    \midrule
    Interactions    & 9,994,420 & 40,415 & 95,269  & 33,210 & 62,059 \\
    Users   & 8,882     & 6,064  & 7,074   & 5,764 & 6,225 \\
    Items   & 193,635   & 9,637  & 22,372  & 11,595 & 12,900\\
    Artists & 29,847    & 5,176  & 8,582   & 5,526 & 5,398 \\
    \bottomrule
    \end{tabular}
\end{table}

\subsubsection{Datasets}
We use two modern MRS datasets for our experiments: Music4All-Onion (\textbf{M4A-Onion}) \cite{moscati2022music4all} and \textbf{Yambda-50m}~\cite{ploshkin2025yambda}. Both contain user-track interaction logs, track metadata (including artist mappings), and audio content for cold track representation. For M4A-Onion, 30-second raw audio previews are available via Music4All \cite{santana2020music4all}. We process these with MuQ-MuLan \cite{zhu2025muq} to generate 512-dimensional audio content vectors, selecting this model for its strong performance in MRS contexts \cite{Tamm_2024}. For Yambda, raw audio is not available, so we instead use the pre-calculated 128-dimensional audio embeddings provided with the dataset, which were generated by a proprietary CLMR-style \cite{spijkervet2021contrastive} model.

\subsubsection{Data Splitting}
Our experiments focus on finding relevant new music for users via item cold-start top-$k$ recommendation, i.e.\ on the retrieval stage of the MRS pipeline. To this end, we process the datasets by converting all interaction logs into unique user-track pairs. We employ a global time split to align with the production cold-start setting and prevent data leakage \cite{meng2020exploring,ji2023critical}, and apply 5-core filtering on training users and items to reduce interaction noise. Moreover, in the cold test set we include only users and artists that were present in the training set. Since the two datasets differ in their time periods and numbers of items available in a cold-start scenario, we apply slightly different splitting strategies.

\textbf{M4A-Onion} has a wide period of time available, but a relatively small number of new items added per month, necessitating a wide test interval to gather more cold items. We select one year of training data from 2017-09-01 to 2018-08-31 and use interactions for items that first appear in the next three months (2018-09-01 to 2018-12-01) as test data, choosing these dates to maximize the number of cold item interactions. We construct our cold validation set (for hyperparameter tuning) by selecting all items that first appeared in the last training month (2018-08-01 to 2018-09-01), excluding them from the training data. This approach aligns the characteristics of the validation set with those of the test set while avoiding making predictions more than three months into the future.

\textbf{Yambda} has only 300 days of data, but a much larger item population with many new tracks appearing each month. We therefore use the last four weeks of data for cold evaluation, dividing the new items in this period into 30/70 splits for the cold validation and cold test sets; all interactions before then are used for training. We omit listening events if the user listened to less than 20\% of the track. 

\paragraph{\textbf{Artist-Aware Evaluation}} \label{sec:aa-eval} Since we use artists to guide our recommendations, we follow the evaluation strategy proposed in \cite{meehan2025artist}, which introduces the notion of known and unknown artists. This is a per-user notion: an artist is \textbf{known} if a user has previously listened to tracks by this artist, and \textbf{unknown} otherwise. The \textbf{Discovery} split contains all test interactions with \textbf{unknown} user-artist pairs, and the \textbf{Exploit} split contains all test interactions with \textbf{known} user-artist pairs. The \textbf{Overall} set refers to the full cold test set, i.e., the union of Discovery and Exploit (rather than the union cold and hot sets, as in other cold-start works). When we evaluate on Discovery, we only include predictions on unknown user-artist pairs; for Exploit, we do the opposite. This allows us to evaluate a model's ability to suggest new tracks by unknown artists (leading to more serendipitous recommendations) separately from new tracks by familiar artists. Dataset split statistics are in Table \ref{tab:data}.

\begin{figure}
    \definecolor{hotitems}{HTML}{4A4A4A}
\definecolor{colditems}{HTML}{3F80EA}
\definecolor{hotartists}{HTML}{FC5353}
\definecolor{coldartists}{HTML}{4A4A4A}
\definecolor{known}{HTML}{EF940D}
\definecolor{unknown}{HTML}{00C090}
\definecolor{labelgray}{HTML}{4A4A4A}

\begin{tikzpicture}[scale=1, every node/.style={scale=1},
    label/.style={font=\fontfamily{phv}\selectfont\footnotesize\bfseries, text=labelgray, anchor=east},
    bartext/.style={font=\fontfamily{phv}\selectfont\scriptsize\bfseries, text=white},
    arrow/.style={-{Stealth[length=1.5mm]}, line width=1pt, rounded corners=2pt},
]

% Dimensions
\def\barwidth{6}  % reduced for column width
\def\barheight{0.45}
\def\cornerrad{2pt}
\def\gap{0.02}  % tiny gap between bar segments
\def\rowgap{0.4}  % vertical gap between rows

% Y positions (computed from rowgap)
\pgfmathsetmacro{\ybot}{0}
\pgfmathsetmacro{\ymid}{\ybot + \barheight + \rowgap}
\pgfmathsetmacro{\ytop}{\ymid + \barheight + \rowgap}

% Data percentages
\def\hotitempct{94}
\def\colditempct{6}
\def\hotartistpct{85}
\def\coldartistpct{15}
\def\knownpct{65}
\def\unknownpct{35}

% Calculate widths
\pgfmathsetmacro{\hotitemwidth}{\barwidth * \hotitempct / 100}
\pgfmathsetmacro{\colditemwidth}{\barwidth * \colditempct / 100}
\pgfmathsetmacro{\hotartistwidth}{\barwidth * \hotartistpct / 100}
\pgfmathsetmacro{\coldartistwidth}{\barwidth * \coldartistpct / 100}
\pgfmathsetmacro{\knownwidth}{\barwidth * \knownpct / 100}
\pgfmathsetmacro{\unknownwidth}{\barwidth * \unknownpct / 100}

% === BAR 1: Test Interactions ===
% Hot Items (left) - rounded on left, square on right
\fill[hotitems, rounded corners=\cornerrad] 
    (0, \ytop) -- (0, \ytop + \barheight) -- 
    (\hotitemwidth - \gap, \ytop + \barheight) -- (\hotitemwidth - \gap, \ytop) -- cycle;
\fill[hotitems] (\hotitemwidth - \gap - 0.1, \ytop) rectangle (\hotitemwidth - \gap, \ytop + \barheight);
\node[bartext] at (\hotitemwidth/2, \ytop + \barheight/2) {Hot Items: \hotitempct\%};

% Cold Items (right) - square on left, rounded on right
\fill[colditems, rounded corners=\cornerrad] 
    (\hotitemwidth + \gap, \ytop) -- (\hotitemwidth + \gap, \ytop + \barheight) -- 
    (\barwidth, \ytop + \barheight) -- (\barwidth, \ytop) -- cycle;
\fill[colditems] (\hotitemwidth + \gap, \ytop) rectangle (\hotitemwidth + \gap + 0.1, \ytop + \barheight);
\node[bartext] at (\hotitemwidth + \colditemwidth/2, \ytop + \barheight/2) {\hspace{1pt}\colditempct\%};

% Label
\node[label] at (-0.2, \ytop + \barheight/2) {Test Interactions};

% === ARROW 1: From cold items to bar 2 ===
% Horizontal part in the middle between bars
\pgfmathsetmacro{\arrowmidy}{(\ytop + \ymid + \barheight) / 2 + 0.1}
\coordinate (arrow1start) at (\hotitemwidth + \colditemwidth/2, \ytop);
\coordinate (arrow1mid1) at (\hotitemwidth + \colditemwidth/2, \arrowmidy);
\coordinate (arrow1mid2) at (\barwidth/2, \arrowmidy);
\coordinate (arrow1final) at (\barwidth/2, \ymid + \barheight + 0.08);

\draw[arrow, colditems] (arrow1start) -- (arrow1mid1) -- (arrow1mid2) -- (arrow1final);

% === BAR 2: Cold Items Breakdown ===
% Hot Artists (left) - rounded on left, square on right
\fill[hotartists, rounded corners=\cornerrad] 
    (0, \ymid) -- (0, \ymid + \barheight) -- 
    (\hotartistwidth - \gap, \ymid + \barheight) -- (\hotartistwidth - \gap, \ymid) -- cycle;
\fill[hotartists] (\hotartistwidth - \gap - 0.1, \ymid) rectangle (\hotartistwidth - \gap, \ymid + \barheight);
\node[bartext] at (\hotartistwidth/2, \ymid + \barheight/2) {Hot Artists: \hotartistpct\%};

% Cold Artists (right) - square on left, rounded on right
\fill[coldartists, rounded corners=\cornerrad] 
    (\hotartistwidth + \gap, \ymid) -- (\hotartistwidth + \gap, \ymid + \barheight) -- 
    (\barwidth, \ymid + \barheight) -- (\barwidth, \ymid) -- cycle;
\fill[coldartists] (\hotartistwidth + \gap, \ymid) rectangle (\hotartistwidth + \gap + 0.1, \ymid + \barheight);
\node[bartext] at (\hotartistwidth + \coldartistwidth/2, \ymid + \barheight/2) {\coldartistpct\%};

% Label
\node[label] at (-0.2, \ymid + \barheight/2) {Cold Items (\colditempct\%)};

% === ARROW 2: From hot artists to bar 3 ===
% Horizontal part in the middle between bars
\pgfmathsetmacro{\arrowmidyb}{(\ymid + \ybot + \barheight) / 2 + 0.1}
\coordinate (arrow2start) at (\hotartistwidth/2, \ymid);
\coordinate (arrow2mid1) at (\hotartistwidth/2, \arrowmidyb);
\coordinate (arrow2mid2) at (\barwidth/2, \arrowmidyb);
\coordinate (arrow2final) at (\barwidth/2, \ybot + \barheight + 0.08);

\draw[arrow, hotartists] (arrow2start) -- (arrow2mid1) -- (arrow2mid2) -- (arrow2final);

% === BAR 3: Hot Artists Breakdown ===
% Known (left) - rounded on left, square on right
\fill[known, rounded corners=\cornerrad] 
    (0, \ybot) -- (0, \ybot + \barheight) -- 
    (\knownwidth - \gap, \ybot + \barheight) -- (\knownwidth - \gap, \ybot) -- cycle;
\fill[known] (\knownwidth - \gap - 0.1, \ybot) rectangle (\knownwidth - \gap, \ybot + \barheight);
\node[bartext] at (\knownwidth/2, \ybot + \barheight/2) {Known: \knownpct\%};

% Unknown (right) - square on left, rounded on right
\fill[unknown, rounded corners=\cornerrad] 
    (\knownwidth + \gap, \ybot) -- (\knownwidth + \gap, \ybot + \barheight) -- 
    (\barwidth, \ybot + \barheight) -- (\barwidth, \ybot) -- cycle;
\fill[unknown] (\knownwidth + \gap, \ybot) rectangle (\knownwidth + \gap + 0.1, \ybot + \barheight);
\node[bartext] at (\knownwidth + \unknownwidth/2, \ybot + \barheight/2) {Unknown: \unknownpct\%};

% Label
\node[label] at (-0.2, \ybot + \barheight/2) {Hot Artists (\hotartistpct\%)};

\end{tikzpicture}
    \caption{Test interactions split on Yambda-50m.}
    \label{fig:split}
\end{figure}
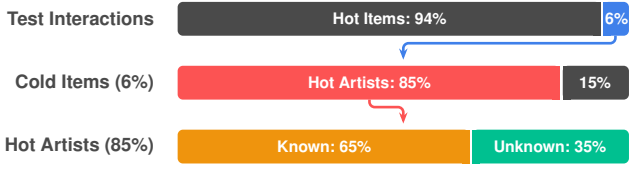

In this paper, we focus on hot artists, since catalog context is unavailable for artists with no previous tracks in the training data. This raises the question: how limiting is this requirement? We cannot make conclusions about general distributions of hot and cold artists; however, as shown in Figure \ref{fig:split}, cold artists comprise only a small fraction of new item interactions in our datasets (15\% on Yambda-50m and 6.5\% on M4A-Onion). ACARec is thus applicable to almost all cold test interactions in these two datasets. 

% \begin{figure}
%     \centering
%     \includegraphics[width=1\linewidth,trim={0.0cm 1.2cm 0.0cm 0}]{img/cold_split_yambda-50m.png}
%     \caption{Test interactions split on Yambda-50m}
%     \label{fig:split}
% \end{figure}

\subsubsection{Metrics}
We measure top-$k$ ranking accuracy on each split with Hit Rate@$k$ (HR@$k$), Recall@$k$ (R@$k$), and Normalized Discounted Cumulative Gain@$k$ (NDCG@$k$ or N@$k$) using standard definitions \cite{Tamm_2021}. We set the ranking cutoff $k=20$, as this is a suitable size for a `New Music For You' playlist on a streaming platform. 

\subsection{Baselines}
\subsubsection{Cold-Start}
\label{sec:cold_start_baselines}
We implement the following cold-start baselines:
\begin{itemize}[leftmargin=5mm]
    \item \textbf{CLCRec}~\cite{CLCREC} applies contrastive learning to align collaborative and content embeddings for improved cold item performance;
    \item \textbf{DeepMusic}~\cite{van2013deep} is trained via  mean-squared error between encoded item content and pre-trained collaborative embeddings;
    \item \textbf{Heater} \cite{HEATER} randomly swaps content and CF vectors during training and transforms content with mixtures-of-experts \cite{shazeer2017outrageously};
    \item \textbf{GAR}~\cite{chen2022generative} implements generative-adversarial training with a content-based generator and pre-trained collaborative model;
    \item \textbf{VBPR}~\cite{he2016vbpr} extends BPR~\cite{BPR} with encoded content features.
\end{itemize}
We implement two variants of each model. The first uses only the audio vector $\mathbf{x}_c$ to represent cold tracks, whereas the second is augmented with artist context. For methods that only learn content projections, we concatenate with the mean CF embedding of the other tracks by the artist (Eq. \ref{eq:artist_mean}), i.e.\ $[\mathbf{x}_h; \overline{\mathbf{e}}_{a}]$; we exclude $h$ from $\overline{\mathbf{e}}_{a}$ during training to simulate cold-start. For CLCRec and VBPR, which train CF embeddings from scratch, we calculate item embeddings as weighted sums between the content outputs and the artist mean in the CF outputs, tuning the balance by Overall validation NDCG. 

\subsubsection{Artist-Based Heuristics}\label{sec:heuristics}
We also evaluate several other approaches for leveraging artist catalogs in the cold item context. \textbf{ArtistMean} represents cold tracks by the artist mean $\overline{\mathbf{e}}_{a}$;  we also test two weighting strategies for this mean. \textbf{ArtistMeanPop} applies weighting based on a track $t$'s popularity (i.e.\ the number of users that listen to $t$ in the training data), on the intuition that user interest in an artist will often be directed towards their most popular tracks. Letting $\mathrm{pop}(t)$ denote $t$'s popularity, we calculate 
\begin{equation}
\overline{\mathbf{e}}^\mathrm{Pop}_{a}=\sum_{j\in\mathcal{H}_a}\frac{\log(\mathrm{pop}(j))}{\sum_{i\in\mathcal{H}_a}\log(\mathrm{pop}(i)) } \mathbf{e}_j.
\end{equation}
However, both this method and ArtistMean do not use any characteristics of the cold track, i.e.\ they will result in the same preference scores for any new track by $a$. We therefore implement another weighted approach, \textbf{ArtistMeanContSim}, that places greater emphasis on tracks with high audio content similarity to the target cold track $t$. Letting $\mathrm{sim}$ represent cosine similarity, we calculate
\begin{equation}
\overline{\mathbf{e}}^\mathrm{ContSim}_{a,t}=\sum_{j\in\mathcal{H}_a}\frac{\exp(\frac{\mathrm{sim}(\mathbf{x}_j, \mathbf{x}_t)}{\tau})}{\sum_{i\in\mathcal{H}_a} \exp(\frac{\mathrm{sim}(\mathbf{x}_i, \mathbf{x}_t)}{\tau})} \mathbf{e}_j,
\end{equation}
i.e.\ the mean weighted by softmax over the content similarities; the temperature $\tau$ is tuned on Overall validation NDCG.

Finally, we implement Personalized Artist Filtering (\textbf{PAF})~\cite{meehan2025artist}, a simple heuristic that suggests only tracks by artists known to the user, ranked by the user's number of listened tracks for that artist.

\subsection{Training Details}
As in other cold-start works~\cite{HEATER, huang_aligning_2023}, we select BPR-based matrix factorization \cite{BPR} as the pre-trained CF model for DeepMusic, Heater, and GAR, as well as ACARec. We tune the embedding size $d_e$ in the range $\{64,128,256,384,512\}$ by hot validation NDCG@50, using a larger cutoff because the hot item set is much larger; the resulting dimensions are 512 for M4A-Onion and 128 for Yambda.

We tune the hyperparameters for all methods by Overall validation NDCG@20. For the baselines, we use parameter search ranges from the original papers. For both DeepMusic and ACARec, the training examples are the hot items, i.e. each epoch makes a single pass through the hot item set $\mathcal{H}$. We train ACARec using Adam \cite{Kingma2014AdamAM} with a learning rate of 0.0005, batch size of 1024 items, and early stopping on Overall validation NDCG@20. We search the number of self-attention and cross-attention heads in $\{2,4,8,16\}$, and the number of artist items sampled during training (i.e.\ the maximum size of $\mathcal{H}'_{a_h}$) in $\{3,5,10,20,30,40,50\}$. For all methods, the reported results are averaged over five runs at the optimal hyperparameters.

\begin{table*}[htbp]
    \centering
    \setlength\extrarowheight{-1pt}
    \small
    \caption{Cold-start results for artist-aware methods. The best model in each metric is \textbf{bolded}, and the second-best is \underline{underlined}. Asterisks (*) indicate statistically significant improvements ($p<0.02$) over the strongest baseline by paired t-test.}
    \begin{tabular}{l| lll|lll|lll}
    \toprule
    & \multicolumn{3}{c|}{Overall} & \multicolumn{3}{c|}{Discovery} & \multicolumn{3}{c}{Exploit} \\
    Model & HR@20 & R@20 & N@20 & HR@20 & R@20 & N@20 & HR@20 & R@20 & N@20 \\
    \midrule
    \multicolumn{10}{c}{\textbf{M4A-Onion}} \\
    \midrule
    PAF                      & 0.5837 & 0.2434 & 0.1909 & —      & —      & —      & 0.7829 & 0.4988 & 0.3233 \\
    ArtistMean               & 0.6990 & 0.3537 & 0.3074 & 0.4357 & 0.2162 & 0.1458 & 0.9070 & 0.6853 & 0.5192 \\
    ArtistMeanPop            & 0.6987 & 0.3563 & 0.3094 & 0.4342 & 0.2145 & 0.1459 & 0.9082 & 0.6884 & 0.5198 \\
    ArtistMeanContSim        & 0.7083 & 0.3552 & 0.3083 & 0.4465 & 0.2150 & 0.1454 & 0.9151 & 0.6890 & 0.5194 \\[2.5pt]
    VBPR + ArtistMean        & 0.6927 & 0.3411 & 0.2904 & 0.4627 & 0.2278 & 0.1553 & 0.9076 & 0.6823 & 0.5117 \\
    Heater + ArtistMean      & 0.7112 & 0.3381 & 0.2903 & 0.4297 & 0.1864 & 0.1239 & 0.9263 & 0.6893 & 0.5135 \\
    CLCRec + ArtistMean      & 0.6589 & 0.2994 & 0.2600 & 0.4600 & 0.2099 & 0.1440 & 0.8914 & 0.6524 & 0.4858 \\
    GAR + ArtistMean         & 0.7269 & 0.3649 & 0.3208 & 0.4863 & 0.2282 & 0.1563 & 0.9220 & 0.6981 & 0.5308 \\
    DeepMusic + ArtistMean   & \textbf{0.7305} & \underline{0.3697} & \underline{0.3239} & \underline{0.5016} & \underline{0.2413} & \underline{0.1667} & \textbf{0.9257} & \underline{0.7029} & \underline{0.5375} \\[2.5pt]
    ACARec (ours)              & \underline{0.7291} & \textbf{0.3733}$^*$ & \textbf{0.3273}$^*$ & \textbf{0.5027} & \textbf{0.2456}$^*$ & \textbf{0.1697}$^*$ & \underline{0.9253} & \textbf{0.7045} & \textbf{0.5410}$^*$ \\
    \midrule
    \multicolumn{10}{c}{\textbf{Yambda-50m}} \\
    \midrule
    PAF                      & 0.1730 & 0.0258 & 0.0207 & —      & —      & —      & 0.1966 & 0.0491 & 0.0300 \\
    ArtistMean               & 0.3329 & 0.0606 & 0.0491 & 0.0800 & 0.0227 & 0.0127 & 0.5317 & 0.1862 & 0.1263 \\
    ArtistMeanPop            & 0.3806 & 0.0749 & 0.0606 & 0.0972 & 0.0283 & 0.0160 & 0.5378 & 0.1875 & 0.1294 \\
    ArtistMeanContSim        & 0.4867 & 0.0982 & 0.0876 & 0.1839 & 0.0537 & 0.0335 & 0.6015 & 0.2012 & 0.1438 \\[2.5pt]
    VBPR + ArtistMean        & 0.3739 & 0.0704 & 0.0608 & 0.1190 & 0.0353 & 0.0210 & 0.5302 & 0.1816 & 0.1305 \\
    Heater + ArtistMean      & 0.5037 & 0.1031 & 0.0912 & 0.1910 & 0.0583 & 0.0329 & 0.6638 & 0.2430 & 0.1788 \\
    CLCRec + ArtistMean      & 0.5439 & 0.1268 & 0.1222 & 0.2414 & 0.0814 & 0.0487 & 0.6645 & 0.2563 & 0.2016 \\
    GAR + ArtistMean         & 0.5777 & 0.1431 & 0.1338 & 0.2700 & 0.0938 & 0.0563 & 0.6788 & 0.2631 & 0.2024 \\
    DeepMusic + ArtistMean   & \underline{0.6256} & \underline{0.1665} & \underline{0.1581} & \underline{0.2834} & \underline{0.0995} & \underline{0.0604} & \underline{0.7298} & \underline{0.3040} & \underline{0.2416} \\[2.5pt]
    ACARec (ours)              & \textbf{0.6498}$^*$ & \textbf{0.1840}$^*$ & \textbf{0.1745}$^*$ & \textbf{0.3356}$^*$ & \textbf{0.1258}$^*$ & \textbf{0.0814}$^*$ & \textbf{0.7431}$^*$ & \textbf{0.3131} & \textbf{0.2492} \\
    \bottomrule
    \end{tabular}
    \label{tab:cold_combined}
\end{table*}

\section{Results}

\subsection{Artist Performance Gain (RQ1)}

\begin{figure}[]
\begin{tabular}{cc}
     \includegraphics[width=38mm,trim={0.7cm 0.2cm 0.4cm 0}]{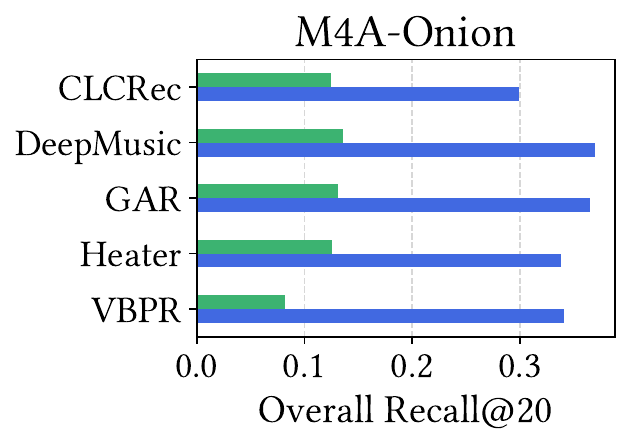}  & \includegraphics[width=38mm,trim={0.4cm 0.2cm 0.7cm 0}]{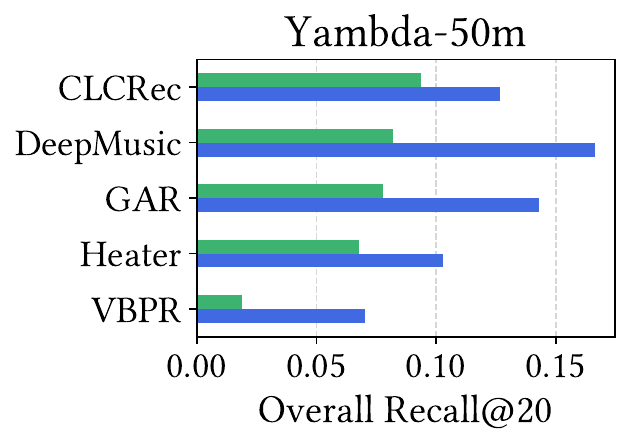} \\
     \includegraphics[width=38mm,trim={0.7cm 0.2cm 0.4cm 0}]{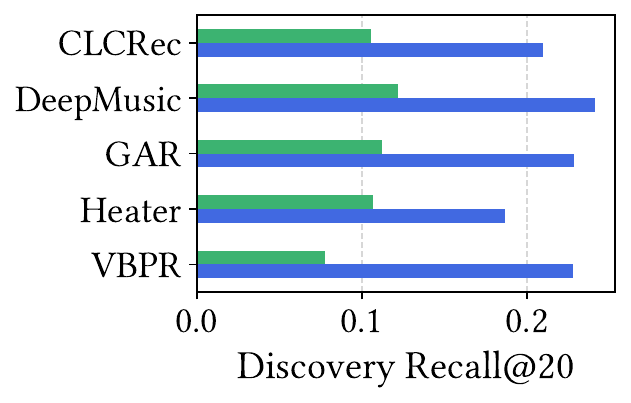}  & \includegraphics[width=38mm,trim={0.4cm 0.2cm 0.7cm 0}]{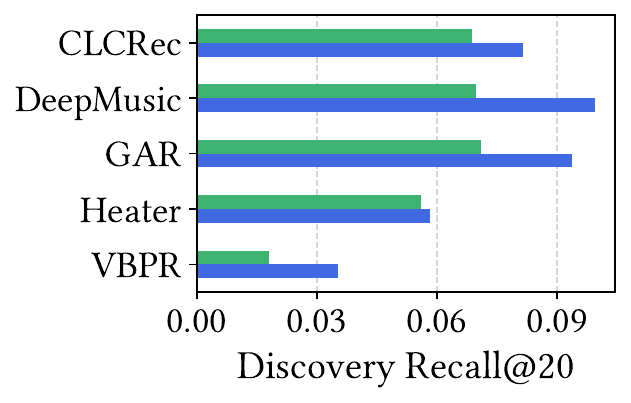}
\end{tabular}
\includegraphics[width=55mm,trim={0 0.5cm 0 0.0cm}]{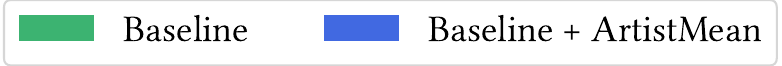}
\caption{Gains from adding ArtistMean into baseline models.} \label{fig:artist_gain}
\end{figure}

We first analyze the impact of leveraging artist signal in cold-start contexts, comparing content-based cold-start methods to their artist-aware modifications (see Section \ref{sec:cold_start_baselines}) in Figure \ref{fig:artist_gain}. Incorporating the ArtistMean significantly increases accuracy, often by up to two or three times in the Overall setting. There are also notable, though less extreme, benefits in Discovery, indicating that artist-related gains are not limited to artists that the user is familiar with. This suggests that user-artist interests are a strong predictive signal for interaction behavior, motivating further exploration of how artist catalogs can be used most effectively in cold track contexts.

Following these findings, we omit content-only baselines from most results below and focus on their artist-aware counterparts.

\subsection{Artist-Aware Cold-Start (RQ2)}

Table \ref{tab:cold_combined} displays cold-start results for artist-aware models. We first observe that metrics are generally higher on M4A-Onion than on Yambda-50m and exhibit less variation between the best and worst models, e.g.\ ArtistMean and its variants have similar accuracy to many of the more complex methods. This is likely explained by the much smaller number of test items in M4A-Onion (1474, versus 22372 in Yambda-50m), as the small number of relevant candidates for each user limits the potential for meaningful performance gains.

\subsubsection{Baselines} Our ArtistMean-based heuristics outperform PAF by a large margin. The Pop and ContSim variants achieve further gains, especially in Yambda, showing that even relatively simple refinement of the artist mean can improve prediction quality; ACARec explicitly integrates this insight into its design.

Of the cold-start baselines augmented with the ArtistMean, GAR and DeepMusic achieve the best results. DeepMusic in particular is the most performant baseline overall. We hypothesize that DeepMusic's training objective, namely the reconstruction of the CF model's item embeddings, is well-suited for augmentation with the ArtistMean inputs, which lie in the same space as the target embeddings. Methods such as Heater and GAR, which attempt to simulate the CF model's ranking behavior, lack this direct connection between inputs and outputs. This insight is further substantiated by noting that DeepMusic consistently sees the largest benefit from the ArtistMean in Figure \ref{fig:artist_gain}, validating our choice of the same reconstruction-based supervision strategy for ACARec.

\begin{figure*}[]
\centering
\includegraphics[width=125mm,trim={0 0.0cm 0 0.0cm}]{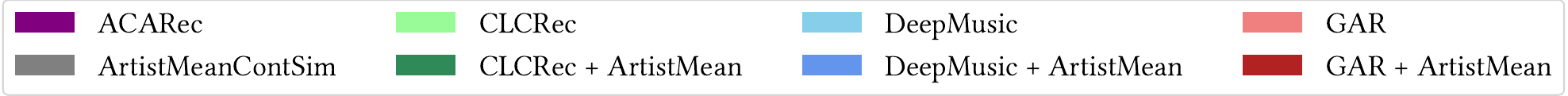}
  \begin{tabular}{ccc}
       \includegraphics[width=54mm,trim={0.1cm 1.1cm 0.1cm 0.0cm}]{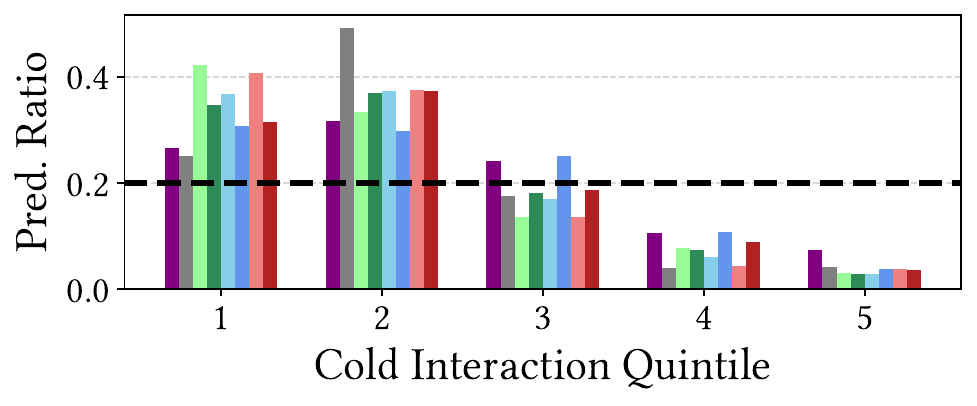} & \includegraphics[width=54mm,trim={0.1cm 1.1cm 0.1cm 0.0cm}]{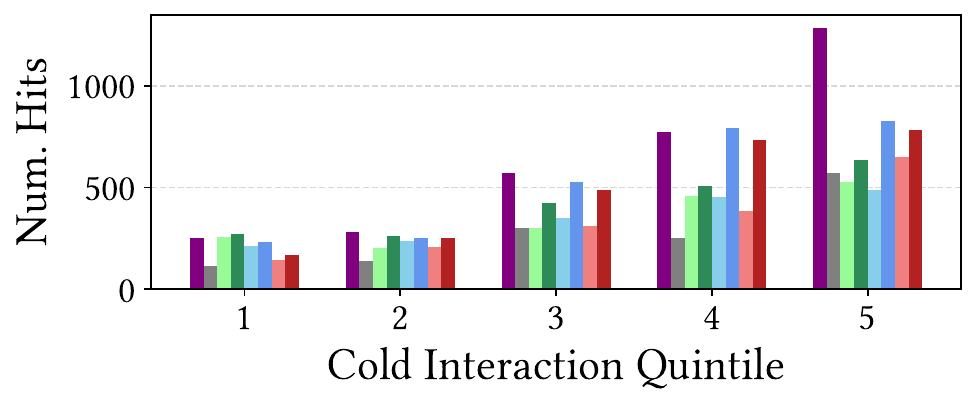} & \includegraphics[width=54mm,trim={0.1cm 1.1cm 0.1cm 0.0cm}]{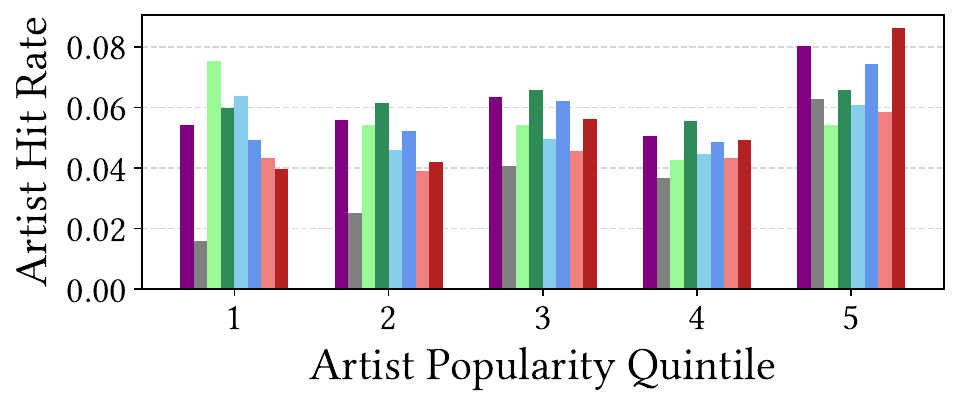} \\[0.3cm] (a) & (b) & (c) \end{tabular} \vspace{-0.33cm}
\caption{Predictive behavior for ACARec and cold-start baselines across interaction and artist popularity quintiles. A higher quintile number indicates that members of that quintile are more popular.}\label{fig:pred_behavior}
\end{figure*}

\subsubsection{ACARec} Our method consistently improves over the baselines, especially in the Overall and Discovery sets. These gains are particularly notable in Yambda, reaching 10.4\% in Overall NDCG@20 and 34.8\% in Discovery NDCG@20. The margins in M4A-Onion are smaller (in part due to the dataset size as discussed above) but still statistically significant in most cases. ACARec's gains in Discovery are especially pertinent given the importance of novelty and diversity to the user experience in MRSs \cite{Content_driven_music_rec,ungruh2024putting}. 

\begin{figure}[]
\centering
\includegraphics[width=\linewidth,trim={0.5cm 0.6cm 0.5cm 0.0cm}]{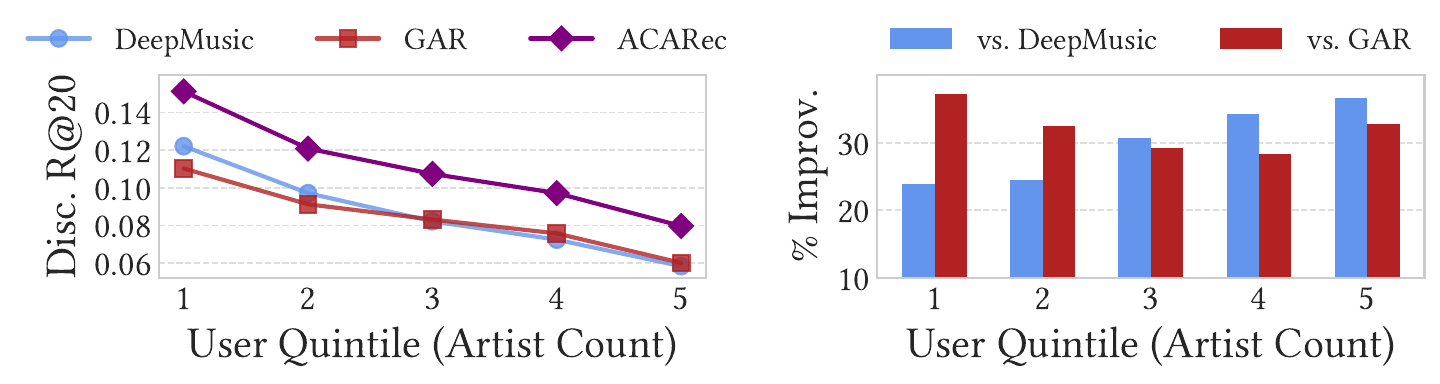}
\caption{Discovery Recall@20 and ACARec's improvement over baselines across user artist count quintiles in Yambda; a higher quintile indicates more interacted artists.}\label{fig:user_improv}
\end{figure}

We examine this further in Figure \ref{fig:user_improv}: we divide the user population into five equally sized groups (quintiles) based on the number of artists they listen to in the training data, and visualize ACARec's accuracy in each group compared to that of GAR and DeepMusic, the two best baselines. We observe that all methods perform worse for users with more listened artists; this aligns with the well-known challenge of capturing diverse user interests in a single embedding~\cite{Guo2021TheSP, cen2020controllable, li2019multi}. This is especially relevant for Discovery, since a newly discovered artist is more likely to differ from the user's previous listening history. However, ACARec's margin of improvement generally grows as the user activity level increases. This is valuable for keeping `power users' on a platform more engaged, and also illustrates that ACARec's elaborate modeling of the artist signal can improve the quality of user personalization.

\subsection{Predictive Behavior (RQ3)}\label{sec:pred}
\subsubsection{Popularity Estimation} A key aim of collaborative interaction modeling is to capture item popularity; this is very challenging in cold-start settings, as, without interaction history, content-based models must predict new item popularity based on content features alone~\cite{meehan2025inherited}. Since artist catalogs provide historical data about potential user interest in new tracks, methods with access to this information should be able to estimate popularity more effectively. 

To test this hypothesis, we analyze model predictive behavior from this popularity perspective; we focus on the Yambda Discovery set, where ACARec has the largest gains over the baselines. We divide tracks into five groups based on their test set interaction count, so that the interactions for all items in each group contains 20\% of the total interactions. For example, the highest popularity group (Interaction Quintile 5), contains only 20 tracks, but these collectively provide 20\% of the Discovery set interactions; the remaining groups contain 82, 430, 4421, and 11901 tracks respectively. We emphasize that, since these items are cold, the model has no data on their popularity, and must predict it based only on its inputs.

In Figure \ref{fig:pred_behavior}(a), we visualize how the proportions of each model's prediction are distributed among the five groups; a model that perfectly captures item popularity will spread its predictions evenly, so that all proportions fall at the reference line at 0.2. We see that all methods are skewed towards the less popular items, illustrating the challenge of recovering popularity in cold-start scenarios. Methods with artist inputs are closer to this line, especially in Groups 3 and 4, but ACARec has the most balanced behavior across the five groups, i.e.\ is able to estimate popularity most accurately. Attending to the most relevant tracks in the artist's catalog allows the model to more effectively predict collaborative behavior for newly added items.

The benefits of this can be seen in the number of hits, or successful predictions, the model makes in each group (Figure \ref{fig:pred_behavior}(b)). ACARec accumulates a large number of Discovery hits in the group with the most popular items; this means that ACARec can not only identify which tracks will be popular, but also the users for which they will be most relevant as an introduction to a new artist. This facilitation of new artist discovery by item content and artist histories alone is highly valuable in music streaming contexts.

\subsubsection{Bias} Although successfully identifying popularity leads to accuracy improvements, it raises concerns about popularity bias~\cite{klimashevskaia2024survey}, i.e.\ that a focus on more popular items may lead to long-tail tracks and artists being under-served. However, we see in Figure \ref{fig:pred_behavior}(b) that ACARec has a similar number of hits to other methods in groups with less popular items, despite allocating slightly fewer predictions to these groups. In Figure \ref{fig:pred_behavior}(c), we evaluate this concern from the artist perspective, dividing artists into five equally sized groups by popularity (i.e.\ listener count in the training data) and measuring the percentage of artists in each group that receive at least one successful prediction (hit rate). Although ACARec's hit rate is slightly higher for more popular artists, the other groups are served roughly equally and in line with other methods. We therefore conclude that our method is able to improve overall artist discovery without sacrificing outcomes for less popular artists; evaluation of additional fairness objectives~\cite{deldjoo2024fairness} will further limit any reinforcement of popularity bias in real-world applications.

\begin{figure}[]
\begin{tabular}{cc}
     \includegraphics[width=38mm,trim={0.7cm 1.1cm 0.2cm 0}]{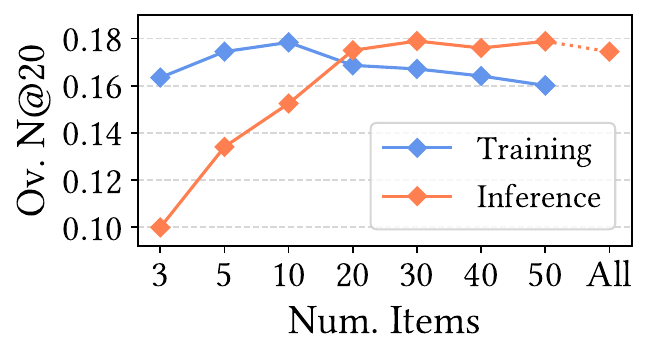}  & \includegraphics[width=38mm,trim={0.7cm 0.7cm 0.2cm 0}]{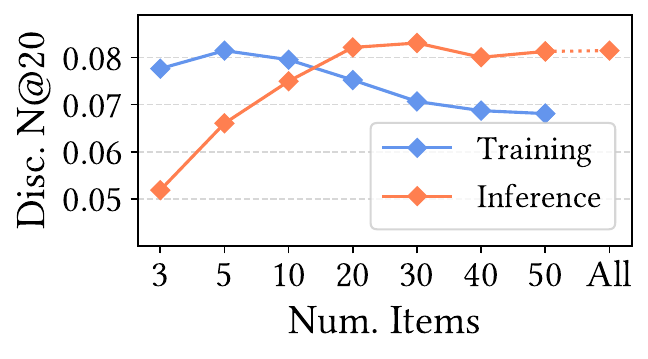} \\
\end{tabular}
\caption{Overall and Discovery NDCG@20 against number of artist items during training and inference on Yambda.} \label{fig:item_num}
\end{figure}

\subsection{Artist Catalog Sampling (RQ4)}\label{sec:sampling_results}
As noted in Section \ref{par:remark}, we sample subsets of artist catalogs during training, and use full artist histories at inference. However, for artists with many tracks, this inference strategy incurs a higher computational cost. We therefore experiment with using subsets of artist tracks during inference. Similar to our ArtistMean baselines, two natural strategies emerge, namely filtering by popularity and by content similarity to the new track. However, the latter still requires querying the entire catalog, so we adopt the former approach and limit artist sets at inference to their top-$n$ tracks by popularity.

In Figure \ref{fig:item_num}, we visualize the impact of the size of the artist subsets for both training (where they are randomly sampled) and inference (where they are filtered by popularity) on the Yambda dataset. We note that the model used for the inference metrics (and for all other reported Yambda results) has subset size five during training, as this configuration achieves the best validation set performance. For the `training' metrics, we use all items in the catalog at inference.

During training, smaller random subsets of five to ten items are optimal, perhaps due to increased stability and training example diversity. In contrast, at least 20 items are needed for inference; however, beyond this 20-item threshold, accuracy remains largely stable. In other words, if querying an artist's entire catalog is impractical, focusing on most-listened tracks provides similar performance. 

We note that track popularity is the simplest criterion for filtering the artist catalog. Extensions of this approach, such as selecting recent tracks, representatives from different albums, or diverse key sets via clustering, are promising directions for future work.

\subsection{Ablation Study (RQ5)}\label{sec:ablation}
We evaluate the contribution of various components of ACARec in Table \ref{tab:ablation}. Both self-attention and late content input meaningfully improve accuracy, although content input has more impact; this shows that, along with being used to query the artist catalog, the content signal adds value in directly predicting the CF embedding.

We also test three alternatives to the GRU for fusing the attention output with the artist prototype. The first, Direct, does not use the prototype and simply reconstructs the embedding from the attention output; we see that this leads to a drop in accuracy. The Residual method (see Section \ref{sec:resid_fusion}), sums the attention output and the artist mean. Although this improves Overall accuracy compared to Direct fusion, it is worse in Discovery, as the artist mean dominates and prevents more nuanced modeling of the cold track. The GLU variant applies a Gated Linear Unit \cite{shazeer2020glu}, commonly used in transformers, as a simpler alternative to the GRU. Including this learnable fusion recovers accuracy in Discovery, but we see that the more sophisticated update gate machinery of the GRU facilitates better preference modeling across all three data splits.

\begin{table}
    \centering
    \setlength\extrarowheight{-1pt}
    \caption{Ablation study on Yambda 50m, reporting NDCG@20 in each data split. Self-Attn. represents the self-attention mechanism over the artist's tracks (Equation \ref{eq:self-attn}) and Cont. Inp. stands for the content concatenation in Equation \ref{eq:cont_inp}.}
    \begin{tabular}{cccccc}
    \toprule
    Self-Attn. & Cont. Inp. & Fusion & Overall & Discov. & Expl. \\
    \midrule
    
    \xmark & \xmark & GRU       & 0.1477 & 0.0669 & 0.2215 \\
    \cmark & \xmark & GRU       & 0.1641 & 0.0721 & 0.2377 \\
    \xmark & \cmark & GRU       & 0.1683 & 0.0767 & 0.2437 \\[2.5pt]
    \cmark & \cmark & Direct    & 0.1541 & 0.0743 & 0.2271 \\
    \cmark & \cmark & Residual  & 0.1580 & 0.0692 & 0.2314  \\
    \cmark & \cmark & GLU  & 0.1672 & 0.0744 & 0.2407  \\[2.5pt]

    \cmark & \cmark & GRU       & 0.1745 & 0.0814 & 0.2492 \\

    \bottomrule
    \end{tabular}
    \label{tab:ablation}
\end{table}

\section{Discussion}
\label{sec:discussion}
% \subsection{Where can this be applied besides music?}
% Remove this for now to save space

\paragraph{\textbf{Limitations}}\label{sec:limitations}
ACARec uses artist catalogs to recommend cold tracks, which naturally limits it to hot artists. As noted in Section \ref{sec:aa-eval}, hot artists account for the vast majority of cold interactions in our datasets; however, the remaining interactions are from new artists who would benefit from promotion to grow their fan base. Since new artists likely also lack other content sources, such as biographies \cite{oramas2017deep}, a default approach would be to fall back to track-only cold-start methods to accumulate more collaborative impressions. Other solutions, e.g.\ generating synthetic artist prototypes by content similarity to existing artists, are potential future work.

We also assume that each track has only one artist; if more than one is listed, we set the first as the only artist of the track. However, in reality, tracks often have multiple artists, e.g.\ collaborations, features, remixes, or covers, where secondary artists can influence user interest. Other artist-artist relationships, such as shared band members, producers, or record labels, could also provide a useful signal, though modeling these may require a more complex graph-based approach. While these more nuanced artist relations might be incorporated into the model, the extent of their effect beyond the gains achieved by our single-artist approach remains to be explored.

\paragraph{\textbf{Future Work}}
ACARec relies on a pretrained hot CF model and uses item content alongside artist context to map into that model's latent space. Since this approach can be applied to any base model, it is an easy-to-adopt practical solution (e.g.\ \cite{briand2024let}). Our catalog attention idea could also be integrated into hot CF models, enriching hot item recommendations with artist context.

As discussed in Section \ref{sec:related_artist}, several works model artist information as text \cite{oramas2017deep, salganik2024larp,Lee2025}. Usually this is an artificial description constructed from metadata, e.g. "The track \texttt{<track name>} by \texttt{<artist name>} on album \texttt{<album name>}" as in LARP \cite{salganik2024larp}. While this captures semantic meaning of artist names, it is unclear how much improvement stems from linking tracks via shared artist identities rather than embedding the names with a text encoder. LLMs provide a convenient and unified but indirect approach to embed artists, whereas ACARec is specialized but attends directly over artist catalogs; comparing their effectiveness warrants further study.

\section{Conclusion}
In this paper, we reframe the track cold-start recommendation problem as a semi-cold artist-aware problem. Our datasets show that 85-93\% of all cold-track interactions belong to artists with previous music available for inferring user preferences. In other words, hot artists are to cold-start recommendation what hot tracks are to standard recommendation: a source of collaborative signal that should be exploited directly. We show that even simple augmentation with an artist mean embedding significantly improves the performance of cold-start track-only models. Building on this foundation, we propose ACARec, an attention-based architecture that generates collaborative embeddings for new tracks by attending over the artist's existing catalog. By learning which catalog tracks are most informative for a given release, ACARec achieves consistent improvements over both naive catalog aggregation and artist-augmented baselines, with substantial gains in artist discovery scenarios. Our analysis shows that ACARec better estimates cold item popularity while maintaining coverage across artists of varying popularity levels. We hope this work encourages more explicit treatment of artist-track relationships in MRS research.

%\clearpage

\bibliographystyle{ACM-Reference-Format}
\bibliography{bibliography}

\end{document}